\documentclass[pre,amsmath,preprintnumbers,nofootinbib,preprint,showpacs]{revtex4}
\usepackage{amsmath,amsfonts,amssymb,bm}
\usepackage{graphicx}

\DeclareMathOperator{\trace}{Tr}

\begin{document}

\title{Multidimensional measures of impulsively driven stochastic systems based on the Kullback-Leibler distance}
\author{Saar Rahav}
\author{Shaul Mukamel}
\address{Department of Chemistry, University of California, Irvine, CA 92697}

\begin{abstract}
By subjecting a dynamical system to a series of short pulses and varying several time delays we can obtain multidimensional characteristic measures of the system. 
Multidimensional Kullback-Leibler response function (KLRF), which are based on the Kullback-Leibler distance between the initial and final states, are defined. We compare the KLRF, which are nonlinear in the probability density, with ordinary response functions (ORF) obtained from the expectation value of a dynamical variable, which are linear. We show that the KLRF encode different level of information regarding the system's dynamics.  For overdamped stochastic dynamics two dimensional KLRF shows a qualitatively different variation with the time delays between pulses, depending on whether the system is initially in a steady state, or in thermal equilibrium.
\end{abstract}


\maketitle

\section{Introduction}

One of the most common ways to investigate the properties of a dynamical system is to study how it responds to  controlled external perturbations. 
The response of a system to a weak perturbing field is related to its equilibrium fluctuations by the celebrated
 fluctuation dissipation relation~\cite{Callen1951,Kubo}. The response provides a direct measure of system dynamics and fluctuations.

In a time-domain response measurement one uses a series of impulsive perturbations (Fig.~\ref{fig1}) and records some property of the system as a function of their time-delays. Impulsive perturbations make it possible to study the free dynamical evolution of the system during the time delays unmasked by the time profile of the perturbing field. Furthermore, the joint dependence on several time delays can be used to separate the contributions of different dynamical pathways. Due to its dependence on multiple time delays this method is termed multidimensional.  

The response of a system is typically measured by the expectation value of some operator~\cite{Callen1951,VanKampen,Kubo}. This is a linear functional of the system probability density (or the Density matrix for quantum systems). Multidimensional response have had considerable success in nonlinear spectroscopy, due to the ability to control and shape optical fields. Applications range from spin dynamics in NMR~\cite{ernstbook}, vibrational dynamics of proteins in infrared systems and electronic energy transfer in photosynthetic complexes as probed by visible pulses~\cite{Mukamel2009,Abramavicius2009}. 
These span a broad range of timescales from milliseconds to femtoseconds.

Interestingly, there exist non-linear functionals of probability densities which have interesting physical interpretations. One such quantity is the Von-Newman entropy $\tilde{S}(\rho)=-\trace \rho \ln \rho$. A related quantum nonlinear measure called the concurrence serves as a measure of quantum entanglement~\cite{Wootters1998,Yu2009}. 
The Kullback-Leibler distance (KLD) or relative entropy, $\trace \rho_0 \ln \rho_0 / \rho$, which compares one probability distribution to another, is a nonlinear measure that had been found useful in many applications. This paper aims at developing multidimensional measures based on the KLD.

 Numerous applications of the KLD differ in the probability distributions involved. The ratio of the probability of a stochastic path and its reverse at a steady state has been connected to a change of entropy~\cite{Lebowitz1999,Maes2003,Andrieux2008,Gomez-Marin2008}. For an externally driven system a similar quantity was found to be related to the work done on the system~\cite{Crooks1998}. As a result, the KLD which compares the path distribution to a distribution of reversed paths is a measure of the lack of reversibility of a thermodynamical process.

For distributions in phase-space (as opposed to path-space), the KLD between the density of a driven system and the density of a reversed process~\cite{Kawai2007}, or the distance between the driven density and the corresponding equilibrium density~\cite{Vaikuntanathan2009} (for the same value of parameters) were shown to be bounded by dissipated work in the process.
 The transfer of information through a stochastic resonance is quantified by the KLD between the probability distributions with and without the external input~\cite{Neiman1996,Goychuk2009}. The ability of neuronal networks to retain information about past events was characterized by the Fisher-information~\cite{Ganguli2008}, which is closely related to the KLD between a distribution and the one obtained from it by a small perturbation.

We shall examine the response of a system to impulsive perturbations which drive it out of a stationary 
(steady state or equilibrium) state. The KLD between the distribution before and after the perturbation 
does not correspond to an entropy, or work. However, since it compares the perturbed and unperturbed densities,
it characterize how ``easy'' it is to drive the system away from its initial state. 
In ordinary response theory, one compared the expectation values of some operator taken over the perturbed 
and unperturbed probability density. This depends on the specific properties of the observed operator. The KLD
is a more robust measure for the effect of the impulsive perturbation on the probability density.
By expanding the KLD in the perturbation strength we obtain a hierarchy of Kullback-Leibler response functions (KLRF).
These differ qualitatively from the hierarchy of ordinary response functions (ORF), since they are nonlinear in the probability density. The KLRF serve as a new type of measures characterizing the dynamics and encoding different information than the ORF.
For example, the second order KLRF, which we connect to the Fisher-information, is found to exhibit qualitatively different dependence on the time delays, depending on whether the system is perturbed out of a steady state or out of thermal equilibrium. This is in contrast to the corresponding ORF. The fluctuation dissipation relation~\cite{Callen1951,Kubo}, which is linear in the density matrix, can also distinguish between systems driven out-of-equilibrium and out of a steady state. The KLD offer a different window into this aspect.

The structure of the paper is as follows. In Sec.~\ref{formalsec} we describe the multidimensional measures and present the two heirarchies of ORF and KLRF response functions. These are then calculated using a formal perturbation theory in the coupling strength to the external perturbation. In Sec.~\ref{overdampedsec} we show that for systems undergoing overdamped stochastic dynamics the non-linear KLRF are naturally described using a combination of the stochastic dynamics and its dual dynamics.
In Sec.~\ref{mastersec} we extend  the results of Sec.~\ref{overdampedsec} to discrete Markovian systems with a finite number of states. Our results are discussed in Sec.~\ref{discsec}.

\section{Multidimensional measures for nonlinear response based on the Kullback-Leibler distance}
\label{formalsec}

\begin{figure}[h]
 \includegraphics[scale=0.85]{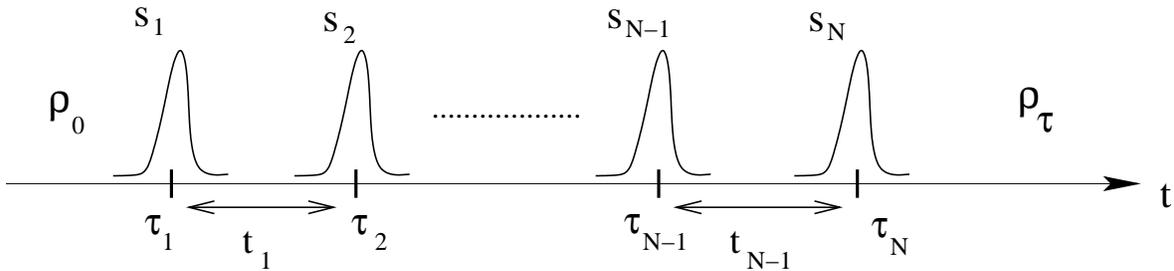}
\caption{Heuristic schematic of the process studied. A system is prepared at an initial probability distribution $\rho_0$ and subjected to impulsive perturbations. The $k$th pulse is centered at $\tau_k$ and its strength is denoted by $s_k$. $t_j=\tau_{j+1}-\tau_j$ are the time intervals between successive pulses. \label{fig1}}
\end{figure}

We consider a system initially at a stationary state (either equilibrium or a steady state), which is perturbed by a series of short pulses, as depicted in Fig.~\ref{fig1}. 
The probability distribution describing the driven system at time $\tau$, $\rho (\tau)$, depends parametrically on $s_i$, the strength of the $i$th pulse, as well as on the time differences between the pulses $t_i \equiv \tau_{i+1}-\tau_i$ with $\tau_{n+1}=\tau$.

Ordinary response theory focuses on the expectation value of some observable
\begin{equation}
\label{avgA}
 \left< A \right>(t) = \trace \left[ \rho (t) A \right],
\end{equation}
and its dependence on the parameters $s_i,\tau_i$. The lowest ordinary response functions (ORF) are
\begin{eqnarray}
 {\cal R}_i^{(1)} (\tau-\tau_i) & \equiv & \left. \frac{\partial \left< A \right>}{\partial s_i} \right|_{{\bf s}=0},\label{defr2} \\
{\cal R}_{ij}^{(2)}(\tau-\tau_i,\tau_i-\tau_j) & \equiv & \left. \frac{\partial^2 \left< A \right>}{\partial s_i \partial s_j} \right|_{{\bf s}=0}, \label{defr3} \\
{\cal R}_{ijk}^{(3)}(\tau-\tau_i,\tau_i-\tau_j,\tau_j-\tau_k) & \equiv & \left. \frac{\partial^3 \left< A \right>}{\partial s_i \partial s_j \partial s_k} \right|_{{\bf s}=0}, \label{defr4}
\end{eqnarray}
and so forth. The time differences in Eqs. (\ref{defr2})-(\ref{defr4}) can be expanded in terms of the time delays between the pulses, $\tau-\tau_i=\sum_{k=i}^n t_k$. 
 ${\cal R}^{(j)}$ are used to investigate various properties of the unperturbed dynamics, such as the existence of excited modes, and the relaxation back to a steady state.

 Here, we focus on different, but closely related quantity. Instead of studying an expectation value of an observable, we focus on a quantity that compares the perturbed and unperturbed probability distributions. The KLD, also known as the relative entropy, is defined as
\begin{equation}
 {\cal D} \left( \rho_0 || \rho \right) \equiv \trace \rho_0 \ln \frac{\rho_0}{\rho} = \left< \ln \frac{\rho_0}{\rho}\right>_0. \label{defkld}
\end{equation}
The KLD vanishes when the two distributions are equal ${\cal D} \left( \rho || \rho \right)=0$, and is positive otherwise, ${\cal D} \left( \rho^\prime || \rho \right) > 0$ for $\rho^\prime (x) \ne \rho (x)$~\cite{Coverbook}.  Note that the KLD is not a true distance since ${\cal D} \left( \rho^\prime || \rho \right) \ne {\cal D} \left( \rho || \rho^\prime \right)$ and it does not satisfy the triangle inequality.

The KLD measures the dissimilarity between two distributions. It had found many applications in the field of information theory~\cite{Coverbook}. For instance, the mutual information between two random variables $x,y$ is ${\cal D} \left( P(x,y)|| P(x) P(y)\right)$, where $P(x,y)$ is the joint distribution while $P(x)$ ($P(y)$) is the marginal distribution of $x$ ($y$).
In the present application the Kullback-Leibler distance is a measure for the deviation of the system from its initial state.

In a manner similar to the definition of the ORF, we define a KLRF hierarchy by taking derivatives of the KLD with respect to pulse strengths, and displaying them with respect to the time delays
\begin{eqnarray}
 {\cal Q}^{(1)}_i (\tau-\tau_i) & \equiv & \left. \frac{\partial {\cal D}}{\partial s_i} \right|_{{\bf s}=0}  =  \left< \left.\frac{\partial {\cal }}{\partial s_i} \ln \frac{\rho_0}{\rho} \right|_{{\bf s}=0} \right>_0,  \\
 {\cal Q}^{(2)}_{ij} (\tau-\tau_i,\tau_i-\tau_j) & \equiv & \left. \frac{\partial^2 {\cal D}}{\partial s_i \partial s_j} \right|_{{\bf s}=0}  =  \left< \left.\frac{\partial^2 {\cal }}{\partial s_i \partial s_j}  \ln \frac{\rho_0}{\rho} \right|_{{\bf s}=0} \right>_0, \label{second} \\
{\cal Q}^{(3)}_{ijk} (\tau-\tau_i,\tau_i-\tau_j,\tau_j-\tau_k)& \equiv & \left. \frac{\partial^3 {\cal D}}{\partial s_i \partial s_j \partial s_k} \right|_{{\bf s}=0}  =  \left< \left.\frac{\partial^3 {\cal }}{\partial s_i \partial s_j \partial s_k}  \ln \frac{\rho_0}{\rho} \right|_{{\bf s}=0} \right>_0. \label{qijk}
\end{eqnarray}
All the derivatives are calculated at ${\bf s}=0$, and we have used the relation $\lim_{{\bf s} \rightarrow 0 } \rho(\tau) = \rho_0$. This is also true for all other $s$ derivatives in the following. To keep the notation simple we will not state this explicitly. Higher order KLRF are defined similarly. It is important to note that the ORF are linear in $\rho$ whereas the KLD are nonlinear. We thus expect the KLD to carry qualitatively different information about the dynamics.

The second derivative (\ref{second}), known as the Fisher Memory (or information) matrix, plays an important role in information theory, since the Cra\'mer-`Rao inequality means that it is a measure of the minimum error in estimating the value of a parameter of a distribution~\cite{Coverbook}. The Fisher information has been used recently to analyze the survival of information in stochastic networks~\cite{Ganguli2008}.

Conservation of probability implies that the first KLRF vanishes
\begin{equation}
 {\cal Q}^{(1)}_i (\tau-\tau_i) = - \int dx \rho_0 \frac{\partial}{\partial s_i} \ln \rho = - \int dx \frac{\partial}{\partial s_i} \rho =0.
\end{equation}

The second derivative, the Fisher memory matrix, is given by
\begin{equation}
 {\cal Q}^{(2)}_{ij} (\tau-\tau_i,\tau_i-\tau_j)= \int dx \rho_0^{-1} \frac{\partial \rho}{\partial s_i} \frac{\partial \rho}{\partial s_j} = \left< \frac{\partial \ln \rho}{\partial s_i}  \frac{\partial \ln \rho}{\partial s_j} \right>_0. \label{jvialn}
\end{equation}
A straightforward calculation allows to recast the third order derivative of ${\cal D}$ in terms of  products of lower order  derivatives
\begin{multline}
 {\cal Q}^{(3)}_{ijk} (\tau-\tau_i,\tau_i-\tau_j,\tau_j-\tau_k)= \\ \left< \frac{\partial \ln \rho}{\partial s_i} \frac{\partial^2 \ln \rho }{\partial s_j \partial s_k}\right>_0 + \left< \frac{\partial \ln \rho}{\partial s_j} \frac{\partial^2 \ln \rho }{\partial s_i \partial s_k}\right>_0 + \left< \frac{\partial \ln \rho}{\partial s_k} \frac{\partial^2 \ln \rho }{\partial s_i \partial s_j}\right>_0 + \left< \frac{\partial \ln \rho}{\partial s_i} \frac{\partial \ln \rho}{\partial s_j} \frac{\partial \ln \rho}{\partial s_k}\right>_0. \label{mvialn}
\end{multline}
In what follows the derivatives will be calculated perturbatively in $s_j$. It is important to note that the $N$'th derivative of $\ln \rho$ has contributions from interaction with at most $N$ pulses. The contribution from the linear component, which interacts with $N$ pulses, has the same structure of the perturbation theory for observables (which is also linear). However, since $\ln \rho$ is a non linear function of $\rho$ the $N$'th derivative contains a non-linear contribution which is a product of lower order contributions for $\rho$.
The KLRF encode qualitatively new information about the system dynamics in comparison to the ORF.

The time evolution of the probability distribution is given by
\begin{equation}
 \frac{\partial \rho}{\partial \tau} = - \hat{\cal L}(\tau) \rho.
\end{equation}
This formal equation is quite general, and can describe either Hamiltonian (Unitary) or stochastic dynamics where the operator $\hat{\cal L}$ will accordingly be the Liouville, or the Fokker-Planck operator.

For a system subjected to a time dependent weak perturbation we can write
\begin{equation}
 \hat{\cal L} (\tau) = \hat{\cal L}_0 + \hat{\cal L}^\prime (\tau),
\end{equation}
where we assume that the unperturbed system is time independent and is intially in a steady state, $\rho_0$, so that $\hat{\cal L}_0 \rho_0 =0$. We consider an impulsive perturbation of the form
\begin{equation}
\label{lprime}
 \hat{\cal L}^\prime (\tau) = -\sum_{i=1}^n s_i \delta (\tau-\tau_i) \hat{\cal L}_{A},
\end{equation}
where $\hat{\cal L}_A$ describes the action of a pulse on the probability distribution, and $s_i$ is the overall strength of the $i$th pulse.

Using these definitions, the state of the system at time $\tau$ can be expanded as a power series in the number of interactions with the pulses
\begin{eqnarray}
\label{pseries}
 \rho (\tau)  & = & \rho_0 + \delta \rho_1 + \delta \rho_2 + \cdots \\ & = & \rho_0+ \sum_{i=1}^n s_i {\cal S}^{(1)}(x;\tau-\tau_i) + \sum_{i=1}^n \sum_{j=1}^i s_i s_j \left(1-\frac{\delta_{ij}}{2} \right) {\cal S}^{(2)} (x; \tau-\tau_i,\tau_i-\tau_j) + \cdots \nonumber
\end{eqnarray}
The partial corrections for the density, ${\cal S}^{(j)}$, appearing in Eq. (\ref{pseries}), contain all the information necessary for computing both the KLRF and ORF. They are given by
\begin{eqnarray}
 {\cal S}^{(1)} (x;t) & \equiv & \int dx_1 dx_2  {\cal U}_0 (x,x_2; t) {\cal L}_A (x_2,x_1) \rho_0 (x_1), \label{defs1} \\
 {\cal S}^{(2)} (x; t^\prime,t) & \equiv & \int dx_1 \cdots dx_4  {\cal U}_0 (x,x_4; t^\prime) {\cal L}_A (x_4,x_3) {\cal U}_0 (x_3,x_2; t) {\cal L}_A (x_2,x_1) \rho_0 (x_1). \label{defs2}
\end{eqnarray}
Here $\hat{\cal U}_0 (\tau-\tau^\prime) = \exp \left[ - (\tau-\tau^\prime) \hat{\cal L}_0 \right]$ is the free propagator of the unperturbed system. Conservation of probability requires that $\int dx \delta \rho_i (x)$, which, in turn, means that $\int dx {\cal S}^{(i)} = 0$.

Eqs. (\ref{pseries})-(\ref{defs2}) can be used to calculate the logarithmic derivatives, which then determine the KLRFs. We will only need the first two logarithmic derivatives, which are given by 
\begin{equation}
 \frac{\partial \ln \rho}{\partial s_i} = \frac{1}{\rho_0} \frac{\partial \delta \rho_1}{\partial s_i} =  \frac{1}{\rho_0 (x)} {\cal S}^{(1)}(x;\tau-\tau_i), \label{dlnr1}
\end{equation}
and
\begin{eqnarray}
 \frac{\partial \ln \rho}{\partial s_i \partial s_j} & = & \frac{1}{\rho_0} \frac{\partial^2 \delta \rho_2}{\partial s_i \partial s_j} - \frac{1}{\rho_0^2} \frac{\partial \delta \rho_1}{\partial s_i} \frac{\partial \delta \rho_1}{\partial s_j} \nonumber \\
& = & \frac{1}{\rho_0 (x)} {\cal S}^{(2)} (x;\tau-\tau_i,\tau_i-\tau_j) - \frac{1}{\rho_0^2(x)} {\cal S}^{(1)}(x;\tau-\tau_i) {\cal S}^{(1)}(x;\tau-\tau_j). \label{dlnr2}
\end{eqnarray}

To calculate the ORF, we substitute
Eq. (\ref{pseries}) in Eq. (\ref{avgA}), resulting in
\begin{equation}
 \left<A \right>_\tau = \left<A \right>_0 + \sum_{n=1}^\infty \int_{\tau_0}^\tau d \tau_n \int_{\tau_0}^{\tau_n} d \tau_{n-1} \cdots \int_{\tau_0}^{\tau_2} d \tau_1 s(\tau_n) \cdots s(\tau_1) {\cal R}^{(n)} (\tau-\tau_n,\tau_n-\tau_{n-1}, \cdots , \tau_2-\tau_1).
\end{equation}

It is interesting to compare the expressions of the KLRF the ORF. We Calculate ${\cal Q}^{(j)}$ and ${\cal R}^{(j)}$ for $j$ perturbing pulses. To leading order, we find
\begin{eqnarray}
{\cal Q}^{(1)}_1 (t_1) & = & 0, \nonumber \\
 {\cal R}^{(1)} (t_1) & = & \int dx A (x) {\cal S}^{(1)} (x;t_1). \label{formalr1}
\end{eqnarray}
At the next order, we compare the Fisher information to the second order ORF,
\begin{equation}
\label{fisher2}
 {\cal Q}^{(2)}_{12} (t_2,t_1)=\int dx \rho_0^{-1} (x) {\cal S}^{(1)} (x;t_1+t_2) {\cal S}^{(1)} (x;t_2),
\end{equation}
and
\begin{equation}
\label{formalr2}
 {\cal R}^{(2)} (t_2,t_1) = \int dx A (x) {\cal S}^{(2)} (x;t_2,t_1).
\end{equation}
The non-diagonal elements of ${\cal Q}^{(2)}_{ij}$ depend on the two delay times.
Expressions for the third order response functions are given in App.~\ref{thirdorderformal}.

Both ${\cal R}^{(j)}$ and ${\cal Q}^{(j)}$ depend on the same set of $j$ time intervals with some important differences.
 ${\cal Q}^{(1)}$ vanishes, while the linear response ${\cal R}^{(1)}$ does not. ${\cal Q}^{(2)}$ and ${\cal R}^{(2)}$ have a different structure: ${\cal R}^{(2)}$ can be calculated from the second order correction to the density (or  ${\cal S}^{(2)}$) while ${\cal Q}^{(2)}$ is determined from a product of ${\cal S}^{(1)}$'s describing the first order interaction with different pulses. This difference reflects the non-linear dependence of the KLRF on $\rho$, and also applies to higher orders.

A comment is now in order regarding our choice of the KLD (\ref{defkld}). We have chosen to use ${\cal D}\left( \rho_0 || \rho \right)$ as the measure for the effect of the perturbations.  ${\cal D}\left( \rho || \rho_0 \right)$ would have been equally suitable. However, as discussed in App.~\ref{hamiltonianapp}, the leading order of both KLDs in the strength of the perturbation, i.e. their Fisher informations, coincide. Therefore all the following results pertaining to the Fisher information would hold for either choice.

\section{Application to overdamped stochastic dynamics}
\label{overdampedsec}

 In the following we use the formal results of Sec.~\ref{formalsec} to calculate the leading order ORF and KLRF for a system undergoing overdamped stochastic dynamics. We show that the Fisher information is related to a forward-backward stochastic process. The backward part is driven by the $\rho_0$-dual process, which will be simply referred to as the dual in what follows. The Fisher information is found to exhibit qualitatively different properties for systems perturbed from equilibrium, or from a steady state. We also use the eigenfunctions and eigenvalues of the dynamics to derive explicit expressions for several low order ORF and KLRF. 

In stochastic dynamics the probability density plays the role of a reduced density matrix, which depends on a few collective coordinates. In this reduced description the entropy $-\trace \rho \ln \rho$ typically increases with time. This should be contrasted with a description which includes all the degrees of freedom, where the dynamics is unitary and the entropy does not change in time. For completeness unitary dynamics is discussed in  App.~\ref{hamiltonianapp}.

\subsection{Fisher information for systems perturbed out of equilibrium vs steady-state}

The Fisher information can be represented in terms of the dual stochastic dynamics. This interesting property  reflects its non-linear dependence on $\rho$. We examine a stochastic dynamics of several variables $x_j$, given by
\begin{equation}
\label{stochasticoriginal}
 \frac{x_i}{dt} = F_i ({\bf x}) + \xi_i (t;{\bf x}).
\end{equation}
Here we use the Ito stochastic calculus.~\footnote{The Ito and Stratonovich calculus offer two different recipes of interpreting Eq. (\ref{stochasticoriginal}). Both methods are equally viable as long as they are used in a consistent manner. Details can be found is Ref.~\cite{Gardiner}.} The noise terms are assumed to be Gaussian with
\begin{equation}
 \left<\xi_i (t) \xi_j (t^\prime)\right> = G_{ij} \delta (t-t^\prime).
\end{equation}
with $G({\bf x})$ a symmetric positive definite matrix. While for many systems this matrix does not depend on the coordinate, $G_{ij}=\frac{2}{\gamma} k_B T \delta_{ij}$, this assumption will not be used in what follows.

Equation (\ref{stochasticoriginal}) is equivalent to the Fokker-Planck equation
\begin{equation}
 \frac{\partial \rho}{\partial t} = - \sum_i \frac{\partial}{\partial x_i} F_i \rho + \sum_{ij} \frac{1}{2} \frac{\partial^2}{\partial x_i \partial x_j} G_{ij} \rho = - \hat{\cal L}_0 \rho.
\end{equation}
 In what follows we present the dual dynamics, which can be loosely thought as the time  reversed dynamics: it have the same steady state, but with reversed steady state current. We consider the current density
\begin{equation}
 J_i=F_i \rho - \sum_j \frac{1}{2} \frac{\partial}{\partial x_j} G_{ij} \rho.
\end{equation}
The Fokker-Planck equation can be written in terms of the current,
\begin{equation}
 \frac{\partial \rho}{\partial t} = - \sum_i \frac{\partial J_i}{\partial x_i}.
\end{equation}

The steady-state is the solution of
\begin{equation}
 \hat{\cal L}_0 \rho_0 =0.
\end{equation}
We write 
\begin{equation}
 \rho_0 \equiv e^{- \phi_0},
\end{equation}
which defines $\phi_0$. For systems at equilibrium $\phi_0$ is simply the potential. However, this is not the case for general steady states. The steady state current can be written as
\begin{equation}
 J_i^{(0)} = F_i \rho_0 - \sum_j \frac{1}{2} \frac{\partial}{\partial x_j} G_{ij} \rho_0  =\frac{1}{2} \sum_{j} G_{ij} \left( \sum_k 2 G^{-1}_{jk} F_k -\sum_{kl} G^{-1}_{jk} \frac{\partial G_{kl}}{\partial x_l} + \frac{\partial \phi_0}{\partial x_j}\right) \rho_0.  
\end{equation}
After some algebra, the generator of the stochastic dynamics can be written in terms of the steady state density and currents~\cite{Chernyak2006}
\begin{equation}
 \hat{\cal L}_0 \rho =  \sum_i \frac{\partial}{\partial x_i} e^{\phi_0} J_i^{(0)} \rho - \frac{1}{2} \sum_{ij} \frac{\partial}{\partial x_i} G_{ij} e^{-\phi_0} \frac{\partial}{\partial x_j} e^{\phi_0} \rho.
\end{equation}

The dual dynamics is given by
\begin{equation}
 \frac{\partial \rho}{\partial t} = {\cal L}^d \rho,
\end{equation}
with
\begin{equation}
 \hat{\cal L}^d = \rho_0 \hat{\cal L}^\dagger_0 \rho_0^{-1}.
\end{equation}
A straightforward calculation gives
\begin{equation}
 \hat{\cal L}^d \rho =  \sum_i \frac{\partial}{\partial x_i} F_i^d \rho - \sum_{ij} \frac{1}{2} \frac{\partial^2} {\partial x_i \partial x_j} G_{ij} \rho
\end{equation}
with
\begin{equation}
 F_i^d = - F_i + \sum_j e^{\phi_0} \frac{\partial}{\partial x_j} G_{ij} e^{-\phi_0}. 
\end{equation}
It is a simple matter to verify that the dual dynamics has the same steady state as the original one, but the steady state currents have opposite signs. It can be simulated by integrating the Ito stochastic equation
\begin{equation}
 \frac{d x_i}{d t} = F_i^d ({\bf x})+ \xi_i (t;{\bf x}).
\end{equation}
 The dual dynamics reverses the non conservative forces in the system. This relates the joint probability to go from one place to another in the original dynamics to the joint probability of the reversed sequence of events in the dual dynamics~\cite{Chernyak2006}
\begin{equation}
\label{conddual}
 P_0 ({\bf x}^\prime, t_1 |{\bf x}) \rho_0 ({\bf x}) = P^d ({\bf x},t_1 | {\bf x}^\prime) \rho_0 ({\bf x}^\prime).
\end{equation}
 The left hand side of  Eq. (\ref{conddual}) is the joint steady state probability to first the system at $x$, and  at $x^\prime$ after a time $t_1$. The right hand side is the joint probability of the reversed sequence of events, but for a modified dynamics. When this modified dynamics is the dual these joint probabilities become equal.

We next turn to discuss the system's response to a series of impulsive perturbation. We assume that the perturbation is of the form
\begin{equation}
\label{LA}
 \hat{\cal L}_A \rho= \frac{1}{\gamma}\sum_i  \frac{\partial}{\partial x_i} \frac{\partial A}{\partial x_i} \rho,
\end{equation}
with $A(x)$ as a potential field perturbing the system.
Using Eq. (\ref{defs1}) we have
\begin{equation}
 S_1 ({\bf x}_1;t_1) = \int d{\bf x}_0 P_0 ({\bf x}_1,t_1|{\bf x}_0) \frac{1}{\gamma} B({\bf x}_0) \rho_0 ({\bf x}_0),
\end{equation}
where 
\begin{equation}
 B({\bf x}) \equiv \sum_i \frac{\partial^2 A}{\partial x_i^2} - \frac{\partial A}{\partial x_i} \frac{\partial \phi_0}{\partial x_i}.
\end{equation}
With the help of equation (\ref{conddual}), we obtain
\begin{equation}
 S_1 ({\bf x}_1,t_1) = \int d{\bf x}_0 \frac{1}{\gamma} B({\bf x}_0) P^d ({\bf x}_0,t_1|{\bf x}_1) \rho_0 ({\bf x}_1).
\end{equation}

We now have all the tools needed to compare the ORF and KLRF for overdamped stochastic dynamics. The leading order response function is given by
\begin{equation}
 {\cal R}_1 (t_1) = \frac{1}{\gamma} \int d{\bf x}_1 d {\bf x}_0 A({\bf x_1}) P_0 ({\bf x}_1 , t_1 | {\bf x}_0) B({\bf x}_0) \rho_0 ({\bf x}_0),
\end{equation}
while ${\cal Q}^{(1)}=0$. At the next order, we have
\begin{equation}
 {\cal Q}_{12}^{(2)} (t_2,t_1) = \frac{1}{\gamma^2} \int d{\bf x}_2 d{\bf x}_1 d {\bf x}_0 B({\bf x}_2) P^d ({\bf x}_2,t_2 | {\bf x}_1) P_0 ({\bf x}_1,t_1+t_2|{\bf x}_0) B({\bf x}_0) \rho_0 ({\bf x}_0), \label{Q12usingdual}
\end{equation}
and
\begin{equation}
 {\cal R}_2 (t_2,t_1) = \frac{1}{\gamma^2} \sum_i \int d {\bf x}_2 d{\bf x}_1 d {\bf x}_0 A({\bf x_2})  P_0 ({\bf x}_2 , t_2 | {\bf x}_1) \frac{\partial}{\partial x_{1i}} \left[ \frac{\partial A}{\partial x_{1i}}  P_0 ({\bf x}_1 , t_1 | {\bf x}_0) \right] B({\bf x}_0) \rho_0 ({\bf x}_0). \label{R2deriv}
\end{equation}

Some insight into the structure of different response functions can be gained by representing them as ensemble averages over stochastic trajectories.
The first order response function can be simulated directly using stochastic trajectories of the original dynamics. The appearance of a derivative of the conditional probability complicates the direct simulation of ${\cal R}_2$. It may be possible to circumvent this difficulty using the finite field method, where one combined simulations with and without a finite, but small perturbation~\cite{Mukamel1996}.
 ${\cal Q}^{(2)}_{12}$  can be simulated with trajectories which follow the original dynamics for time $t_1+t_2$ and then the dual dynamics for time $t_2$. 

Systems at equilibrium are self dual, allowing to substitute  $\int d {\bf x}_1 P^d ({\bf x}_2,t_2|{\bf x}_1) P_0 ({\bf x}_1,t_1+t_2|{\bf x}_0)) = P_0 ({\bf x}_2,t_1+2t_2|{\bf x}_0)$in Eq. (\ref{Q12usingdual}) . As a result the Fisher information only depends on a single time variable $t_1+ 2 t_2$. This is in contrast to systems which are perturbed from a nonequilibrium steady state, whose Fisher information is a two dimensional function of $t_1$ and $t_2$. The Fisher information is therefore qualitatively different for systems which are perturbed out of a steady state, or out of an equilibrium state.

For the self-dual case ${\cal Q}^{(2)}_{12}$ has the same structure as ${\cal R}^{(1)}$, up to a replacement of $A(x)$ with $B(x)$. However, in the general, non self-dual case, the structure of ${\cal Q}_{12}^{(2)}$ is manifestly different than that of ${\cal R}^{(1)}$ and ${\cal R}^{(2)}$.

\subsection{Eigenfunction expansion of the Fisher information}

An alternative approach for the calculation of the Fisher information, as well as higher order KLRF, uses eigenfunction expansions for the density $\rho$. It will be sufficient to examine a simple one dimensional model where $\hat{\cal L}_0$ is a Fokker-Planck operator and the perturbation is given by Eq. (\ref{LA}).

The propagation of the unperturbed system can be described in terms of the eigenvalues and eigenfunctions of $\hat{\cal L}_0$.
The right eigenfunctions satisfy
\begin{equation}
 \hat{\cal L}_0 \rho_n = \Lambda (n) \rho_n.
\end{equation}
Similarly, the left eigenfunctions, $q_n$, satisfy $q_n {\cal L}_0 = \Lambda (n) q_n$. It is assumed that the right and left eigenfunctions constitute a byorthogonal system, that is
\begin{equation}
 \int dx q_n (x) \rho_m (x) = \delta_{nm}.
\end{equation}
Any probability density can be expanded in terms of right eigenfunctions,
\begin{equation}
 \rho(x) = \sum_n C_n \rho_n (x),
\end{equation}
with
\begin{equation}
 C_n = \int dx q_n(x) \rho(x).
\end{equation}

We consider systems perturbed out of thermal equilibrium, with a probability density $\rho_0 (x)$. In this case the left eigenvalues are simply related to the right eigenvalues. For variable $x$ which is even with respect to time reversal, this relation takes the form~\cite{Gardiner}
\begin{equation}
\label{LRrelation}
 \rho_n (x) = q_n (x) \rho_0 (x).
\end{equation}
As a result, one can use only the left eigenfunctions, which in this case satisfy the orthogonality condition
\begin{equation}
\label{orthleft}
 \int dx q_n (x) q_m (x) \rho_0 (x)=\delta_{nm}.
\end{equation}
We note that $q_0(x)=1$ and $\Lambda(0)=0$ correspond to the equilibrium distribution.

Our goal is to calculate the response of the system to a perturbation with a coordinate dependent operator $A(x)$. Several types of integrals appear repeatedly in the calculation, and it will be convenient to introduce an appropriate notation.
One comes from the need to decompose the probability distribution into eigenstates after each interaction with a pulse
\begin{equation}
 {\cal B}_{mn} \equiv \int dx q_m(x) \frac{\partial}{\partial x} \left[\frac{\partial A}{\partial x} q_n(x) \rho_0(x) \right] = -\int dx \frac{\partial q_m}{\partial x} \frac{\partial A}{\partial x} q_n(x) \rho_0(x),\label{defB}
\end{equation}
where integration by parts was used in the second equality. (We assume that $\rho_0(x)$ falls of fast enough to eliminate boundary terms.)
The calculation of the response also involves an evaluation of the average of an observable, which, in the current setting, leads to integrals of the form
\begin{equation}
 {\cal C}_{An} \equiv \int dx A(x) q_n(x) \rho_0 (x). \label{defC}
\end{equation}

It is straightforward to calculate ${\cal S}^{(1)}$ and ${\cal S}^{(2)}$ with the help of Eq. (\ref{defB}). We find
\begin{equation}
 {\cal S}_1 (x;t_1)= \sum_n {\cal B}_{n0} e^{-\Lambda (n) t_1} q_n (x) \rho_0 (x),
\end{equation}
and
\begin{equation}
 {\cal S}^{(2)} (x;t_2,t_1) = \sum_{mn} {\cal B}_{mn} {\cal B}_{n0} e^{-\Lambda (n) t_1 -\Lambda (m) t_2} q_m (x) \rho_0 (x).
\end{equation}
We now calculate the first few ORF and KLRF. The first order response functions is
\begin{equation}
 \label{R1stoch}
{\cal R}_1 (t_1) = \sum_n {\cal C}_{An} {\cal B}_{n0} e^{-\Lambda (n) t_1}. 
\end{equation}
Similarly, the second order response function is given by
\begin{equation}
 \label{R2stoch}
{\cal R}_2 (t_2,t_1) = \sum_{mn} {\cal C}_{Am} {\cal B}_{mn} {\cal B}_{n0} e^{-\Lambda(n) t_1 - \Lambda (m) t_2}.
\end{equation}
${\cal R}^{(2)}$ should be compared to the KLRF of the same order, namely the Fisher information
\begin{equation}
 {\cal Q}^{(2)}_{12} (t_2,t_1) = \sum_n {\cal B}_{n0}^2 e^{- \Lambda (n) \left[t_1 + 2 t_2 \right]},
\end{equation}
which is calculated with the help of Eqs. (\ref{fisher2}) and (\ref{orthleft}).
Again, the Fisher information of systems perturbed out of equilibrium depends only on $t_1+2 t_2$. 

For a system initially in a steady state the Fisher information is 
\begin{equation}
 {\cal Q}^{(2)}_{12} (t_2,t_1) = \sum_{nm} {\cal B}_{n0} {\cal B}_{m0} e^{- \Lambda (n) \left[t_1 +  t_2 \right] -  \Lambda (m) t_2} \int dx \rho_0^{-1} (x) \rho_n (x) \rho_m (x). \label{geneigenQ}
\end{equation}
However, since the relation Eq. (\ref{LRrelation})  does not hold in this case, the integral in Eq. (\ref{geneigenQ})  does not vanish for $n \ne m$, and the Fisher information becomes a two dimensional function of $t_1$ and $t_2$.

 ${\cal Q}^{(3)}_{123}$ and ${\cal R}^{(3)}$ are calculated in App.~\ref{thirdordereigen}, where it is shown that ${\cal Q}^{(3)}_{123}$ is a three dimensional function of all its time variables. This qualitatively different signature of systems initially at steady state vs equilibrium is unique to the Fisher information, due to its quadratic dependence on $\delta \rho$. It does not apply to higher order quantities, such as ${\cal Q}^{(3)}$.

The eigenfunctions for a simple example, of an harmonic oscillator with an exponential perturbation, are presented in App.~\ref{exampleosc}.

\section{Master equations, dual dynamics, and the Fisher information for discrete systems}
\label{mastersec}

The description of the KLRF in terms of a combination of the regular stochastic dynamics and its dual holds also for Markovian systems with a finite number of states.
Below we derive a simple expression for the Fisher information of a stochastic jump process in terms of the dual dynamics of the original process. 

\subsection{Dual dynamics of discrete Markovian systems}

Consider a system with a finite number of states, undergoing a Markovian stochastic jump process described by the master equation
\begin{equation}
 \label{master}
\dot{\bf P} = {\bf \cal R} {\bf P}.
\end{equation}
${\bf P}$ is a vector of probabilities to find the system in its states and ${\bf \cal R}$ is the transition rate matrix~\cite{VanKampen}. Its off diagonal elements are positive, ${\cal R}_{ij} >0$ for $i \ne j$, and express the rate of transitions from state $j$ to $i$, given that the system is at $j$. The diagonal elements satisfy $\sum_i {\cal R}_{ii}=0$. We further assume that there exist a unique steady state, ${\bf P}^{(s)}$, satisfying
\begin{equation}
 {\bf \cal R} {\bf P}^{(s)} =0,
\end{equation}
and that at this steady state there is a non vanishing probability to find the system at each of the states $i$, $P_i^{(s)} \ne 0$.
The master equation is one of the simplest models for irreversible stochastic dynamics. Below we briefly describe some of its relevant properties, such as the backward equation, and its dual dynamics.

One can define an evolution operator 
\begin{equation}
 {\bf P}(t)= {\cal U}_t {\bf P} (0),
\end{equation}
which satisfy the equation of motion 
\begin{equation}
 \frac{\partial}{\partial t} {\cal U}_t = {\cal R} {\cal U}_t,
\end{equation}
with the initial condition ${\cal U}_0 =1$.

For this model a dynamical variable ${\bf A}$ is a vector with a value corresponding to each state of the system. Its expectation value is given by
\begin{equation}
 \left<{\bf A}\right>_t \equiv {\bf A} \cdot {\bf P} (t) = \sum_i A_i P_i (t).
\end{equation}
Instead of calculating this average by propagating ${\bf P}$, one can define a time dependent dynamical variables, ${\bf A} (t)$, such that $\left<{\bf A}\right>_t \equiv {\bf A} (t) \cdot {\bf P} (0)$. This is the analogue of the Scr\"odinger and Heisenberg pictures in Quantum Mechanics. The equation of motion for ${\bf A} (t)$  is easily shown to be
\begin{equation}
 \frac{\partial}{\partial t} {\bf A} (t) = {\cal R}^\dagger {\bf A}(t).
\end{equation}
This equation is known as the backward equation, which turns to be related to the dual dynamics. (The backward equation is also written as a function of an {\em initial} rather than a final time.)  ${\cal R}$ and ${\cal R}^\dagger$
have the same eigenvalues. However, the roles of the right and left eigenvectors are interchanged and  ${\bf A} (t)$ decays to a uniform vector with at long times.

Let us now define a diagonal matrix ${\bf \Pi}$, using $\Pi_{ii} = P^{(s)}_i$. The dual evolution is then defined as
\begin{equation}
 {\cal R}^d \equiv \Pi {\cal R}^\dagger \Pi^{-1}.
\end{equation}
The fact that $\Pi$ is built from the steady state of ${\cal R}$, ${\bf P}^s$, guaranties that ${\cal R}^d$ is a physically reasonable rate matrix, that is, that is satisfies $\sum_i {\cal R}^d_{ij}=0$. The dual dynamics describes a physically allowed process which has the same steady state as the original process it was derived from. However, at this steady state the dual currents have opposite signs compared to the steady state currents of the original dynamics.
 A process is self-dual, that is, ${\cal R}^d = {\cal R}$, if and only if it
satisfies detailed balance. Self duality is therefore related to being in thermal equilibrium. 

\subsection{Perturbation theory and the Fisher information}

Consider a system subjected to several impulsive perturbations
\begin{equation}
 {\cal R} (\tau) = {\cal R}_0 + \sum_\alpha s_{\alpha} {\cal R}_A \delta (\tau-\tau_\alpha).
\end{equation}
Where ${\cal R}_A$ corresponds to some physical perturbation $A$. Here the free evolution is given by
\begin{equation}
 {\cal U}^{0}_t = e^{{\cal R}_0 t}.
\end{equation}
Similarly the evolution during an impulsive perturbation can be described by
\begin{equation}
\label{impulse}
 {\bf P} (\tau_\alpha+\epsilon)=e^{s_{\alpha} {\cal R}_A} {\bf P} (\tau_\alpha-\epsilon),
\end{equation}
where $\epsilon$ is arbitrarily small.

We expand the exponent in Eq. (\ref{impulse}), and collect all terms of the same order in $s_\alpha$.  The expression for the Fisher information, following Eq. (\ref{fisher2}) is
\begin{equation}
 {\cal Q}^{(2)}_{12} (t_2,t_1) = \sum_i \frac{1}{P^{s}_i} \left({\cal U}^0_{t_2} {\cal R}_A {\bf P}^s \right)_i \left({\cal U}^0_{t_1+t_2} {\cal R}_A {\bf P}^s \right)_i.
\end{equation}
This expression can be simplified by writing one of the propagators in terms of the dual process, using the relation
\begin{equation}
 e^{{\cal R}_0^\dagger t} \Pi^{-1} = \Pi^{-1} e^{{\cal R}^d t}.
\end{equation}
After some algebraic manipulations, we find
\begin{equation}
\label{fisherdual}
 {\cal Q}^{(2)}_{12} (t_2,t_1) = \sum_i \left( {\cal R}_A^d e^{{\cal R}^d t_2} e^{{\cal R}_0 (t_1+t_2)} {\cal R}_A {\bf P}^s \right)_i.
\end{equation}
In Eq. (\ref{fisherdual}) we have defined ${\cal R}_A^d \equiv \Pi {\cal R}_A^\dagger \Pi^{-1}$. ${\cal R}_A^d$ is not the dual of ${\cal R}_A$ since $\Pi$ is not composed of the eigenvector of ${\cal R}_A$
which corresponds to a vanishing eigenvalues. ($\Pi$ is composed of the eigenvalue of ${\cal R}_0$.)

Equation (\ref{fisherdual}) is the analogue of Eq. (\ref{Q12usingdual}) for discrete systems. It demonstrates that for this model the fisher information can be simulated using a combination of the ordinary process and its dual. However one must also include a (possibly artificial) dual perturbation, ${\cal R}_A^d$, which can nevertheless be computed from the known physical one. As before, Equation (\ref{fisherdual}) shows that the Fisher information, for self-dual systems, is a function of $t_1+2 t_2$. 

As a side remark, the model considered here has only even 'degrees of freedom' under time reversal. More general models may also include odd degrees of freedom, such as momenta. In that case one may speculate that there would be anti-self-dual systems whose dual turns out to be the time reversed dynamics. This would lead to a Fisher information depending on $t_1$ alone, as is the case for Unitary dynamics.

\section{Discussion}
\label{discsec}

In this work we have studied a system driven from its steady state by a sequence of impulsive perturbations. We have defined a new set of measures for the response of the system to the perturbation, the KLRF, which are given by the series expansion of the Kullback-Leibler distance between the perturbed and unperturbed probability distributions.

At each order the KLRF and ORF depend on the same time differences between the pulses. However, there are important differences stemming from the nonlinear dependence of the KLRF on $\rho$. The expression for the KLRF, for instance Eqs. (\ref{fisher2}) and (\ref{Q123}) reveal quantities which can be simulated using several trajectories which end at the same point. We have shown that a simpler, but equivalent description exists. It uses the dual dynamics which allows to ``run some of the trajectories backward''. This description is especially appealing for the Fisher information. Instead of viewing the Fisher information as composed over sum of pair of trajectories joined at their end point one can view it as an average of contributions of a single forward-backward trajectory.

Another difference between the KLRF and the ORF has to do with the appearance of derivatives of conditional probabilities, which for {\em deterministic} systems would correspond to groups of very close trajectories. These first appear in ${\cal S}^{(2)}$, see for instance Eqs. (\ref{R2deriv}) for ${\cal R}^{(2)}$. The nonlinear character of the KLRF means that such terms appear in comparatively higher order of the perturbation theory. For example, ${\cal S}^{(2)}$ contributes to ${\cal R}^{(2)}$ but not to the Fisher information. Instead it first contributes to ${\cal Q}^{(3)}_{123}$. 

 We have demonstrated that the Fisher information behaves in a qualitatively different way depending on whether the system is perturbed from equilibrium or from an out-of-equilibrium steady state. For the classically stochastic systems considered here we have seen that in the former case ${\cal Q}^{(2)}_{12}(t_1,t_2)= f(t_1+2t_2)$. That is, the Fisher information is a one dimensional function of the two time delays. This qualitative difference results from the self-duality of equilibrium dynamics, which is another expression of the principle of detailed balance.
${\cal Q}_{123}^{(3)}$ does not show a similar reduction of dimension [see Eq. (\ref{eigenQ3})]. This property is special to the Fisher information.

We have focused on the properties of the Fisher information for overdamped stochastic dynamics. Do other types of systems exhibit similar behavior? In App.~\ref{hamiltonianapp} we consider deterministic Hamiltonian systems. In that case the unitary dynamics results in a Fisher information which depends only on the time $t_1$. It will be of interest to study how (not overdamped) stochastic systems bridge between the overdamped and unitary limits.

We showed that the KLRF can serve as a useful measure characterizing the system's dynamics. They encode information which differs from the information encoded in the ORF. This is demonstrated by the ability of the Fisher information to distinguish between systems perturbed out of equilibrium or out of a non-equilibrium steady state. We expect other useful properties of the KLRF to be revealed by further studies.

\section*{Acknowledgments}

One of us (S. M.) wishes to thank Eran Mukamel for most useful discussions.
The support of the National Science Foundation (Grant No. CHE-0745892) and the National Institutes of Health (Grant No. GM-59230) is gratefully acknowledged.

\begin{appendix}

\section{Unitary dynamics}
\label{hamiltonianapp}

For completeness, in this Appendix we discuss the application of nonlinear response theory to deterministic Hamiltonian systems.  We use simple examples to clarify the relation between the response of a system and the Kullback-Leibler distance. The Hamiltonian of the system is assumed to be of the form ${\cal H}={\cal H}_0-s(\tau) A$, where $A$ is the perturbation. We start with some general comments stemming from the fact that the dynamics in phase space is an incompressible flow.

\subsection{General considerations}

The state of a classical system is described by its phase space density $\rho$. It is important to note that this is the full probability density, which includes information on all the degrees of freedom.
Let us denote the propagator of the classical trajectories by $h$, so that ${\bf x}(\tau)=h_\tau ({\bf x}(0))$ where
${\bf x}$ denotes a phase space point (all coordinates and momenta). Similarly, the propagator for the probability distribution is denoted by ${\cal U} (\tau)$. Liouville theorem tells us that phase space volumes do not change in time. As a result, there is a simple relation between the phase space density at different times,
\begin{equation}
\label{transport}
 \rho ({\bf x},\tau)={\cal U}(\tau) \rho ({\bf x},0) = \rho (h_{-\tau} ({\bf x}),0)=\rho ({\bf x}(0),0).
\end{equation}
The density is simply transported with the dynamics in phase space. The density at $h_\tau ({\bf x})$ at time $\tau$ is equal to the density at ${\bf x}$ at time $0$.

This property of the phase space dynamics have an interesting consequence. Any integral whose integrand depends locally on $\rho$ alone is time independent, since the values of the integrand are just transported around by the dynamics. An interesting example is the entropy function
\begin{equation}
 \tilde{\cal S} (\tau) \equiv - \int d {\bf x} \rho({\bf x},\tau) \ln \rho ({\bf x}, \tau).
\end{equation}
Let us change the integration variable to ${\bf x}_0 = h_{-\tau} ({\bf x})$, which is just the phase space point which would flow to ${\bf x}$ after a time $\tau$. Liouville theorem assures us that the Jacobian of the transformation is unity and therefore
\begin{equation}
\label{constent}
 \tilde{\cal S}(\tau) = - \int d {\bf x}_0 \rho ({\bf x}_0,0) \ln \rho ({\bf x}_0,0) \equiv \tilde{\cal S}_0,
\end{equation}
where we have used Eq. (\ref{transport}). It is clear that this entropy does not depend on time.

Eq. (\ref{constent}) is a result of the unitary evolution in phase space. It connects with all the intricate problems related to the emergence of macroscopic irreversibility out of microscopic reversible dynamics. This deep problem is beyond the scope of the current paper. Such problems are circumvented when one uses a reduced probability density, whose dynamics is irreversible to begin with.

Unitary dynamics lead to an interesting result for the reverse Kullback-Leibler distance
\begin{equation}
 {\cal D} (\rho || \rho_0) \equiv \trace \rho \ln \frac{\rho}{\rho_0} = \trace \rho \ln \rho - \trace \rho \ln \rho_0.
\end{equation}
We have seen that the first term is a constant of the motion. Assuming a Hamiltonian system intially in equilibrium $\rho_0 = e^{-\beta {\cal H}_0}/Z$. We have
\begin{equation}
 {\cal D} (\rho || \rho_0) = -\tilde{\cal S}_0 + \trace \rho \beta \left({\cal H}_0 -{\cal F}_0 \right)= \beta \trace \left( \rho-\rho_0\right) {\cal H}_0,
\end{equation}
where ${\cal F}_0 = -k_B T \ln Z$ is the free energy of the initial equilibrium state. ${\cal D}(\rho || \rho_0)$ is therefore linear in $\rho$, and thus it is equivalent to a calculation of response functions! 

The general comments above raise two interesting points. First, the fact that terms such as $\trace \rho \ln \rho$ are constant means that their expansion in powers of pulse strengths $s_i$ turns out to have only the constant term, all other terms in the expansion must vanish. 

The second point of interest has to do with the relation between the Kullback-Leibler distance and its reverse. Generally, ${\cal D} (\rho||\rho_0) \ne {\cal D} (\rho_0 || \rho)$. We are interested in systems pertubed by a series of pulses and compare the initial and final distributions. In that case we can use unitarity to change variables from the phase space points at the final time to the points at the initial time which are connected to it by the dynamics. This gives
\begin{equation}
 {\cal D} (\rho || \rho_0) = \int d {\bf x} \rho({\bf x},\tau) \ln \frac{\rho ({\bf x},\tau)}{\rho_0 ({\bf x})} = \int d {\bf x}_0 \rho_0 ({\bf x}_0) \ln \frac{\rho_0 ({\bf x}_0)}{\tilde{\rho} ({\bf x}_0, 0)} = {\cal D} (\rho_0 || \tilde{\rho}).
\end{equation}
Here $\rho_0 ({\bf x} (\tau))=\tilde{\rho} ({\bf x}_0,0)$, that is, $\tilde{\rho}$ is the density that would evolve to the equilibrium density under the influence of the pulses. It is not equal to $\rho ({\bf x},\tau)$, which tells us that the distance and its reverse are not equal. 

While in general the distance and its reverse are not equal, when the distance between the distributions is small, its easy to show that their leading order expansion in $\delta \rho = \rho-\rho_0$ is the same. Loosely speaking
\begin{eqnarray}
 {\cal D} (\rho_0 || \rho)  & =  &  \frac{1}{2}\int d {\bf x} \left(\frac{\delta \rho^2}{\rho_0} + O(\delta \rho^3)  \right) \\
{\cal D} (\rho || \rho_0)  & =  &  \frac{1}{2}\int d {\bf x} \left(\frac{\delta \rho^2}{\rho_0} + O(\delta \rho^3)  \right). \\
\end{eqnarray}
We can deduce that the Fisher information, which is determined by this leading order, could be obtained from a calculation of response for systems with unitary evolution. (Since it could also be obtained from the reverse distance.)

Our last general point is also related to unitarity, but has to do with the Fisher information. According to Eq. (\ref{fisher2}) the Fisher information is built from two partial densities. (More accurately, density differences.) These evolve with respect to the same Hamiltonian ${\cal H}_0$ between interactions with the pulses. As a result, the integral in Eq. (\ref{fisher2}) do not depend on $t_2$, for the same reason that caused the entropy $\tilde{\cal S}$ to be time independent. We find that
\begin{equation}
 {\cal Q}_{12}^{(2)} (t_2,t_1) = {\cal Q}_{12}^{(2)} (t_1),
\end{equation}
as a result of the unitarity of the phase space dynamics. Similar independence on the final time interval would also appear for higher order terms in the expansion of the Kullback-Leibler distance.

\subsection{A single Harmonic oscillator}

The simplest system that can serve as an example is a single Harmonic oscillator
\begin{equation}
 {\cal H}_0 = \frac{P^2}{2 M}+ \frac{1}{2} M \Omega^2 Q^2.
\end{equation}
We take the perturbation to be
\begin{equation}
 A(Q)=\alpha e^{-Q/Q_0}.
\end{equation}

For this system it is trivial to solve for the free evolution
\begin{eqnarray}
 Q(\tau) & = & Q(0) \cos \Omega \tau + \frac{P(0)}{M \Omega} \sin \Omega \tau \nonumber \\
P(\tau) & = & P(0) \cos \Omega \tau - \Omega M Q(0) \sin \Omega \tau.
\end{eqnarray}
This relation can be inverted, expressing $P(0),Q(0)$ in terms of $P(\tau),Q(\tau)$
\begin{eqnarray}
 Q(0) & = & Q(\tau) \cos \Omega \tau - \frac{P(\tau)}{M \Omega} \sin \Omega \tau \nonumber \\
P(0) & = & \Omega M Q (\tau) \sin \Omega \tau + P(\tau) \cos \Omega \tau. \label{reverseosc}
\end{eqnarray}
Eq. (\ref{reverseosc}) will be useful when one propagates probability distributions in time.

The equilibrium distribution of the Harmonic oscillator is a Gaussian
\begin{equation}
 \rho_0 = \frac{2 \pi k_B T}{\Omega} e^{-\frac{\beta P^2}{2 M}-\frac{1}{2}\beta M \Omega^2 Q^2}.
\end{equation}
 We also note that
\begin{equation}
\label{tempdrho}
 {\cal L}_A \rho_0= \left\{ A, \rho_0 \right\}=\frac{\beta \alpha}{M Q_0} P e^{-Q/Q_0} \rho_0,
\end{equation}
is the linear order correction for the density just after interaction with one pulse.

To calculate ${\cal S}^{(1)}$ we need to propagate Eq. (\ref{tempdrho}). With the help of Eqs. (\ref{transport}) and (\ref{reverseosc}) we find
\begin{equation}
 \label{s1osc}
{\cal S}^{(1)} (Q,P;t_1) = \frac{\beta \alpha}{M Q_0} \left( \Omega M Q \sin \Omega t_1 + P \cos \Omega t_1 \right) e^{- \frac{1}{Q_0} \left[ Q \cos \Omega t_1 - \frac{P}{M \Omega} \sin \Omega t_1 \right]} \rho_0 (Q,P).
\end{equation}
In the derivation we have used the fact that $\rho_0$ is invariant under the evolution with respect to ${\cal H}_0$.

To calculate ${\cal S}^{(2)}$ we operate with ${\cal L}_A$ on ${\cal S}^{(1)}$, and then propagate the resulting correction for the density for a time interval $t_2$. A straightforward calculation leads to
\begin{eqnarray}
 {\cal S}^{(2)} (Q,P;t_2,t_1) & = & {\cal U} (t_2) {\cal L}_A {\cal S}^{(1)} \nonumber \\
 &  = & - \frac{\beta \alpha^2}{M Q_0^2} \left\{ \cos \Omega t_1 + \left( \Omega M Q \sin \Omega (t_1+t_2) + P \cos \Omega (t_1+t_2)\right) \rule{0pt}{15pt} \right. \nonumber \\ & \times &\left. \left(-\frac{\beta}{M} \left[\Omega M Q \sin \Omega t_2 + P \cos \Omega t_2 \right] + \frac{1}{Q_0 M \Omega} \sin \Omega t_1 \right) \right\}  \\ & \times & \exp \left[-\frac{Q}{Q_0} \left( \cos \Omega t_2 + \cos \Omega (t_1+t_2)\right) +\frac{P}{M \Omega Q_0}\left(\sin \Omega t_2 + \sin \Omega (t_1+t_2) \right)\right] \rho_0 \nonumber
\end{eqnarray}

We now turn to calculate the first two ORF, using Eqs. (\ref{formalr1}) and (\ref{formalr2}). The calculation of ${\cal R}^{(1)}$ is cumbersome but straightforward.
\begin{equation}
 {\cal R}^{(1)} (t_1) = \int dQ dP A(Q) {\cal S}^{(1)} (Q,P;t_1) = - \frac{\alpha ^2}{\Omega M Q_0^2} \sin \Omega t_1 \exp \left[ \frac{1}{\beta M Q_0^2 \Omega^2_0} \left( 1+\cos \Omega t_1\right)\right].
\end{equation}
The calculation of ${\cal R}^{(2)}$ is more involved, and we only include the final result
\begin{multline}
 {\cal R}^{(2)} (t_2,t_1) = \frac{\alpha^3}{M^2 \Omega^2 Q_0^4} \sin \Omega t_2 \left[\sin \Omega t_1 + \sin \Omega (t_1+t_2)\right] \\ \times \exp \left[ \frac{1}{M \beta \Omega^2 Q_0^2} \left\{ \frac{3}{2} +  \cos \Omega t_1 +  \cos \Omega t_2 +  \cos \Omega (t_1+t_2)\right\}\right].
\end{multline}

We would like to compare these response functions to the KLRF, and in particular to the Fisher information, which can be calculated using Eq. (\ref{fisher2}). The calculation can be simplified by using the classical coordinates right after the second interaction with the pulse as integration variables. Due to unitarity, the Fisher information only depends on the time difference $t_1$, see also the discussion in the previous subsection.

We get
\begin{equation}
{\cal Q}_{12}^{(2)} (t_2,t_1) = \frac{\beta \alpha^2 }{M Q_0^2} \left[ \cos \Omega t_1 - \frac{1}{\beta M \Omega^2 Q_0^2}\sin^2 \Omega t_1\right] \exp \left[ \frac{1}{\beta M \Omega^2 Q_0^2} \left( 1+\cos \Omega t_1\right)\right].
\end{equation}
This expression seems similar to the first order response function. One can indeed show that they are related by
\begin{equation}
 {\cal Q}_{12}^{(2)} (t_2,t_1)=\beta \frac{\partial}{\partial t_1} {\cal R}^{(1)} (t_1).
\end{equation}
It will be interesting to check whether this expression could be generalized to any Hamiltonian system (with unitary dynamics).

\section{Third order ORF and KLRF}

In the main text we have calculated the ORF and KLRF to second order.
In the appendix we present the third order quantities, ${\cal R}^{(3)}$, and  ${\cal Q}^{(3)}_{123}$.

\subsection{Perturbation theory}
\label{thirdorderformal}

Following the calculations performed in Sec.~\ref{formalsec}, the third order response functions are given by
\begin{equation}
 {\cal R}^{(3)} (t_3,t_2,t_1)= \int dx A(x) {\cal S}^{(3)} (x;t_3,t_2,t_1),
\end{equation}
and
\begin{multline}
 {\cal Q}^{(3)}_{123} (t_3,t_2,t_1) = \int dx \left\{ \rho_0^{-1} (x) \left[ {\cal S}^{(1)}(x;t_1+t_2+t_3) {\cal S}^{(2)} (x; t_3,t_2) \right. \right. \\ \left. \left. + {\cal S}^{(1)}(x; t_2+t_3) {\cal S}^{(2)}(x;t_3,t_1+t_2)  + {\cal S}^{(1)}(x;t_3) {\cal S}^{(2)}(x; t_3+t_2,t_1)\right] \right. \\ \left. -2 \rho_0^{-2}(x) {\cal S}^{(1)}(x;t_1+t_2+t_3) {\cal S}^{(1)}(x; t_2+t_3) {\cal S}^{(1)}(x; t_3) \right\},\label{Q123}
\end{multline}
where
\begin{multline}
 {\cal S}^{(3)} (x;t_3,t_2,t_1) \equiv \int dx_1 \cdots dx_6 {\cal U}_0 (x,x_6;t_3) {\cal L}_A (x_6,x_5) {\cal U}_0 (x_5,x_4;t_2) {\cal L}_A (x_4,x_3) \\ \times {\cal U}_0 (x_3,x_2;t_1) {\cal L}_A (x_2,x_1) \rho_0 (x_1).
\end{multline}

\subsection{Eigenvalue expansions}
\label{thirdordereigen}

Here we present expressions for the third order ORF and KLRF in term of the eigenvalues and eigenfunctions of the stochastic dynamics.
At high orders the nonlinear character of the contributions to the Kullback-Leibler distance result in integrals with products of several eigenfunctions. Here we will only need the one with three eigenfunctions
\begin{equation}
 {\cal J}_{nml} \equiv \int dx q_n (x) q_m (x) q_l (x) \rho_0 (x). \label{defJ}
\end{equation}

The third order ORF is given by
\begin{equation}
  \label{R3stoch}
{\cal R}^{(3)} (t_3,t_2,t_1) = \sum_{lmn} {\cal C}_{Al}{\cal B}_{lm} {\cal B}_{mn} {\cal B}_{n0} e^{-\Lambda(n) t_1 - \Lambda (m) t_2-\Lambda (l) t_3}.
\end{equation}
This ORF should be compared to the third order KLRF, which is calculated using Eq. (\ref{Q123}).
We find
\begin{multline}
 {\cal Q}^{(3)}_{123} (t_3,t_2,t_1) =  \sum_{mn} {\cal B}_{m0} {\cal B}_{mn} {\cal B}_{n0} \left\{ e^{-\Lambda (n) t_2} e^{-\Lambda (m) \left[ t_1+t_2+2t_3\right]} \right. \\ \left. + e^{-\Lambda(n) \left[ t_1 +t_2 \right]} e^{-\Lambda (m) \left[ t_2+ 2 t_3\right]} + e^{-\Lambda (n) t_1} e^{-\Lambda (m) \left[ t_2 + 2 t_3\right]} \right\}  \\  -  2 \sum_{nml} {\cal B}_{n0} {\cal B}_{m0} {\cal B}_{l0} {\cal J}_{nml} e^{-\Lambda (n) \left[ t_1+t_2 + t_3\right]} e^{-\Lambda (m) \left[ t_2 + t_3\right]} e^{-\Lambda (l) t_3}. \label{eigenQ3}
\end{multline}
The expression for ${\cal Q}^{(3)}_{123}$, presented in Eq. (\ref{eigenQ3}), has three terms in which the orthogonality condition (\ref{orthleft}) has been used, pointing to a reduction of dimension in the time dependence of this specific term. However, the time combinations in these terms are all different. In addition, the fourth term in Eq. (\ref{eigenQ3}) clearly depends on all its time variables. We conclude that
in contract to the Fisher information, the higher order KLRF depend on all their time variables. The reduction of dimension is therefore specific for the Fisher information. It results from the fact that it is built out of a single product of two density corrections.

\section{The overdamped harmonic oscillator}
\label{exampleosc}

In this Appendix we consider a simple example of an
 overdamped harmonic oscillator with a potential $1/2 M \Omega Q^2$, with a perturbing potential $A(Q) = \alpha e^{-Q/Q_0}$. For this system it is possible to write explicit expressions for the eigenvalues and eigenfunctions of the Fokker-Planck operator, as well as to perform several of the integral, defined in Sec.~\ref{overdampedsec}. In this case
\begin{eqnarray}
 \hat{\cal L}_0 \rho & = & - \frac{1}{\gamma} \frac{\partial}{\partial Q} \left[ M \Omega^2 Q \rho + k_B T \frac{\partial}{\partial Q} \rho \right] \\
\hat{\cal L}_A \rho & = & - \frac{\alpha}{Q_0 \gamma} \frac{\partial}{\partial Q} \left[ e^{-Q/Q_0} \rho \right].
\end{eqnarray}
This model has been studied extensively~\cite{Gardiner}. The equilibrium density is given by
\begin{equation}
 \rho_0 (Q) = \sqrt{\frac{M \Omega^2}{2 \pi k_B T}} e^{- \frac{M \Omega^2}{2 k_B T} Q^2}.
\end{equation}
The left eigenfunctions, and the eigenvalues, of this model are
\begin{eqnarray}
 q_n (Q) & = & \sqrt{\frac{1}{2^n n!}} H_n \left( Q \sqrt{\frac{M\Omega^2}{2 k_b T}}\right) \\
\Lambda (n) & = & \frac{M \Omega^2}{\gamma} n.
\end{eqnarray}
Here, $H_n$ are the Hermit polynomials.

The following properties of the Hermit polynomials
\begin{eqnarray}
 \frac{d H_n}{d x} & = & 2 n H_{n-1} (x) \label{Hn1}, \\
H_{n} (x) & = & (-1)^n e^{x^2} \frac{d^n}{d x^n} e^{-x^2}, \label{Hn2}
\end{eqnarray}
together with the exponential form of $A(Q)$, allow to compute some of the integrals defined in Eqs. (\ref{defB}) - (\ref{defJ}) explicitly.

For instance,
\begin{equation}
 {\cal C}_{An} = \alpha \int d Q e^{-Q/Q_0} q_n (Q) \rho_0 (Q) = \alpha \sqrt{\frac{1}{2^n n!}} \int d y e^{- \frac{y}{Q_0} \sqrt{\frac{2 k_B T}{M \Omega^2}}} H_n(y) e^{-y^2}.
\end{equation}
One can substitute Eq. (\ref{Hn2}) for the Hermit polynomial, and use integration by parts. This leads to
\begin{equation}
 {\cal C}_{An} = \alpha (-1)^n \sqrt{\frac{1}{2^n n!}} Q_0^{-n} \left( \frac{2 k_B T}{M \Omega^2}\right)^{n/2} \int dy e^{-y^2} e^{-\frac{y}{Q_0} \sqrt{\frac{2 k_B T}{M \Omega^2}}}.
\end{equation}
This is a Gaussian integral which is easily evaluated to give
\begin{equation}
 {\cal C}_{An} = (-1)^n \alpha \sqrt{\frac{\pi}{2^n n!}} Q_0^{-n}  \left( \frac{2 k_B T}{M \Omega^2}\right)^{n/2} e^{\frac{k_B T}{2 Q_0^2 M \Omega^2}} \label{osccalC}.
\end{equation}
One can also easily calculate the integrals
\begin{equation}
 {\cal B}_{n0} = - \int dQ \frac{\partial q_n}{\partial Q} \frac{\partial A}{\partial Q} \rho_0 (Q) = \frac{2 n}{\sqrt{2^n n!}} \frac{\alpha}{\gamma Q_0} \sqrt{\frac{M \Omega^2}{2 \pi k_B T}} \int dy H_{n-1} (y) e^{-y^2} e^{- \frac{y}{Q_0} \sqrt{\frac{2 k_B T}{M \Omega^2}}},
 \end{equation}
by the same technique. One finds
\begin{equation}
 {\cal B}_{n0} = (-1)^{n-1} \frac{2 n}{\sqrt{2^n n!}} \frac{\alpha}{\gamma Q_0^n} \left( \frac{2 k_B T}{M \Omega^2}\right)^{n/2-1}  e^{\frac{k_B T}{2 Q_0^2 M \Omega^2}}. \label{osccalB0}
\end{equation}

Comparing Eqs. (\ref{osccalC}) and (\ref{osccalB0}), we see that for this model
\begin{equation}
 {\cal B}_{n0} = - n \frac{M \Omega^2}{\gamma k_B T \sqrt{\pi}} {\cal C}_{An}.
\end{equation}
As a result, there is a simple relation between the Fisher information and the first order response function,
\begin{equation}
 {\cal Q}_{12}^{(2)} (t_2,t_1) \propto \left. \frac{\partial {\cal R}_1}{\partial t_1} \right|_{t_1 \rightarrow t_1+2 t_2}.
\end{equation}
This relation is a special result for this model, and is not expected to hold for other systems.

One can also obtain explicit results for ${\cal B}_{nm}$ with non vanishing indices. However, the calculation and the result are quite combersome, and are omitted. We were not able to calculate ${\cal J}_{nml}$ explicitly.

\end{appendix}

\end{document}